\documentclass[nohyperref]{article}
\usepackage[T1]{fontenc}

\usepackage{microtype}
\usepackage{graphicx}
\usepackage{subfigure}
\usepackage{booktabs} 
\usepackage{alphabeta}

\usepackage{hyperref}

\usepackage[accepted]{icml2023}

\usepackage{amsmath}
\usepackage{amssymb}
\usepackage{mathtools}
\usepackage{amsthm}

\usepackage[capitalize,noabbrev]{cleveref}

\theoremstyle{plain}

\theoremstyle{definition}

\theoremstyle{remark}

\usepackage[textsize=tiny]{todonotes}

\icmltitlerunning{PLANTAIN}

\begin{document}

\twocolumn[
\icmltitle{PLANTAIN: Diffusion-inspired Pose Score Minimization for Fast and Accurate Molecular Docking}

\begin{icmlauthorlist}
\icmlauthor{Michael Brocidiacono}{unc}
\icmlauthor{Konstantin I. Popov}{unc}
\icmlauthor{David Ryan Koes}{pitt}
\icmlauthor{Alexander Tropsha}{unc}
\end{icmlauthorlist}

\icmlaffiliation{unc}{Eshelman School of Pharmacy, University of North Carolina at Chapel Hill}
\icmlaffiliation{pitt}{Department of Computational and Systems Biology, University of Pittsburgh}

\icmlcorrespondingauthor{Michael Brocidiacono}{mixarcid@unc.edu}

\icmlkeywords{Machine Learning, ICML}

\vskip 0.3in
]



\printAffiliationsAndNotice{} 

\begin{abstract}
Molecular docking aims to predict the 3D pose of a small molecule in a protein binding site. Traditional docking methods predict ligand poses by minimizing a physics-inspired scoring function. Recently, a diffusion model has been proposed that iteratively refines a ligand pose. We combine these two approaches by training a pose scoring function in a diffusion-inspired manner. In our method, PLANTAIN, a neural network is used to develop a very fast pose scoring function. We parameterize a simple scoring function on the fly and use L-BFGS minimization to optimize an initially random ligand pose. Using rigorous benchmarking practices, we demonstrate that our method achieves state-of-the-art performance while running ten times faster than the next-best method. We release PLANTAIN publicly and hope that it improves the utility of virtual screening workflows.
\end{abstract}

\section{Introduction}

Proteins play many vital roles in the human body, and their functions can be modified by small molecules that bind to them. Designing such ligands for protein targets is an essential task in drug discovery. Molecular docking aims to predict the 3D pose of the ligand in the protein binding site. Docking is commonly used in structure-based virtual screening, wherein a large library of compounds are docked to a protein structure of interest. Their poses are used to score their binding affinity and the highest-scoring compounds are selected for experimental testing \cite{maia_structure-based_2020}. Virtual screening promises to quickly discover binders for a target, but its utility is currently limited due to the low accuracy of docking algorithms \cite{bender_practical_2021}. 

\subsection{Prior Work}
Traditionally, molecular docking has been framed as a problem of energy minimization \cite{trott_autodock_2010, friesner_glide_2004}. The ground-truth pose is considered to be the one that minimizes the potential energy of the protein-ligand complex. Calculating the true energy is intractable due to its quantum mechanical nature; thus, many scoring functions have been developed to approximate it.  Unfortunately, these physics-based approaches need to make many assumptions in order to run quickly. Notably, they generally assume that the protein binding site is inflexible, and only modify the ligand pose. This means the scoring functions are often very sensitive to the specific conformation of the given protein structure.

Recently, several methods have been proposed that use machine learning (ML) to directly predict ligand poses \cite{stark_equibind_2022, lu_tankbind_2022}. Notably, DiffDock \cite{corso_diffdock_2022} is a diffusion mode that predicts a pose by initializing the ligand in a random conformation and iteratively updating the translation, rotation, and torsional angles.

One disadvantage of these ML approaches is that they focus on the problem of whole-protein (or ``blind'') docking, implicitly combining the tasks of binding site prediction and ligand pose prediction. In practice, whole-protein docking makes the problem unnecessarily difficult; in most practical applications, researchers already know the binding site of their protein target and only want compounds that bind to that site \cite{yu_deep_2023}. In this paper, therefore, we focus on the problem of predicting the ligand pose given the protein binding site (``site-specific`` docking).

\begin{figure*}[ht]
\vskip 0.2in
\begin{center}
\centerline{\includegraphics[width=\textwidth]{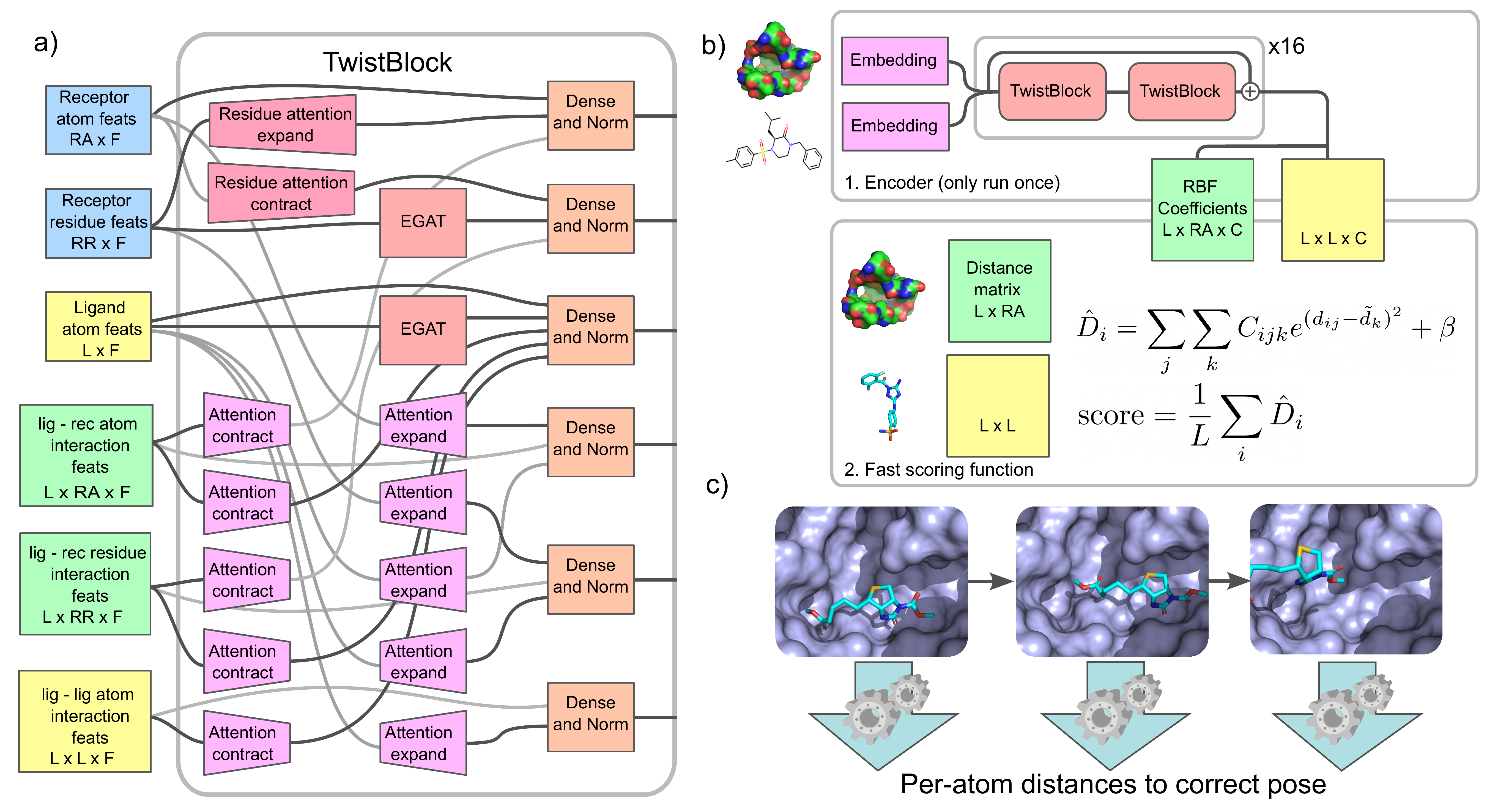}}
\caption{Overview of the PLANTAIN workflow. a) The Twister encoder is composed of 32 TwistBlocks. Each TwistBlock updates the internal representation of the ligand and receptor atom features, receptor residue features, and all the interaction features. b) 1. The encoder uses the protein binding pocket and 2D ligand representation to produce coefficients for the scoring function. The encoder is only run once during inference. 2. The scoring function uses the coefficients from the encoder, along with all the ligand-ligand and ligand-residue inter-atomic distances, to quickly score a proposed ligand pose. c) When training, we add Gaussian noise to the translations, rotations, and torsional angles of the ground-truth ligand pose. We use our model to predict each ligand atom's distance to its correct position.}
\label{overview}
\end{center}
\vskip -0.2in
\end{figure*}

\subsection{A Diffusion-inspired Pose Scoring Function}

We note that both traditional docking methods that rely on energy minimization and the diffusion-based DiffDock share some cursory similarities. Namely, both methods initialize a ligand in a random pose and permute the pose (by translating, rotating, and modifying torsional angles) to give the final prediction. Traditional methods explicitly minimize an energy function, while the diffusion approach uses a neural network to propose new poses directly. Because DiffDock does not explicitly compute any pose score when the pose is generated, it requires a separate confidence model to rank the poses generated by the diffusion model.

We hypothesized that we could train a pose scoring model in a diffusion-inspired manner. When inferring, we then explicitly minimize this score. This approach is conceptually simpler than DiffDock. Additionally, by explicitly training on proteins and ligands from different crystal structures, we hoped that our learned scoring function would be less sensitive to the specific protein conformation than physics-based methods.

We validated this hypothesis by developing PLANTAIN (Predicting LigANd pose wiTh an AI scoring functioN). PLANTAIN uses a novel neural network that, given a protein binding pocket and a \textbf{2D} representation of the ligand, will compute coefficients for a scoring function that can then be quickly computed for a given \textbf{3D} ligand pose. This score function computes the predicted mean absolute error (MAE) of the current pose as compared to the correct pose. We then use this function to minimize the score of the ligand pose using the L-BFGS optimizer \cite{liu_limited_1989}. Using the rigorous benchmarking techniques outlined below, we show that our method improves upon the accuracy of GNINA \cite{mcnutt_gnina_2021}, the next best method, while running significantly faster.

\subsection{The Importance of Rigorous Benchmarking}

Proper benchmarking of docking methods is difficult; it is very easy for hidden biases in the data to artificially inflate reported accuracy. We identify three best practices that should be followed to make docking benchmarks align as closely as possible with actual prospective docking runs. 


First, one should not simply take a ligand out of a crystal structure and attempt to re-dock it to the protein from that structure. Protein residues are flexible and will change conformation upon binding to a particular ligand. Thus, we cannot assume knowledge of the exact locations of the protein pocket residues. The more realistic way to perform benchmarking is to dock a ligand to a \textit{different} crystal structure of the same protein. This procedure is called cross-docking \cite{francoeur_three-dimensional_2020, mcnutt_gnina_2021}.


Second, it is important to ensure the method is tested on proteins unseen in the training set. Previous ML work has frequently used time-based splits to separate train and validation/test data. While such splits ensure that the same protein-ligand complex is not present in both train and test splits, the same proteins can be present in both. ML methods can achieve deceptive performance by memorizing the kinds of interactions made by the protein in question. Additionally, proteins that are very similar to one another should be placed in the same data split. Thus, the most rigorous way to split the data is to first cluster proteins based on sequence (or, better yet, structural) similarity, and place different clusters in different splits \cite{kramer_leave-cluster-out_2010, francoeur_three-dimensional_2020}.


Finally, when docking to a defined binding site, additional bias can be introduced by the definition of the binding pocket. Prior work has often defined the pocket by drawing a box around the ligand with a certain amount of padding. However, the actual pocket might be larger, and we cannot assume knowledge of the specific location within the pocket. Thus, in this paper, we follow \citet{brocidiacono_bigbind_2022} and define the binding site using x-ray crystallographic poses of all known ligands, not just the ligand in question. To our knowledge, this paper is the first work to report docking results using this more rigorous binding pocket definition.

\section{Methods}

\subsection{Model Architecture}

PLANTAIN consists of two parts: (i) a neural network encoder that uses a 3D protein pocket and a 2D ligand graph to produce coefficients, and (ii) a scoring function that uses these coefficients to rapidly evaluate candidate ligand poses. Following Autodock Vina \cite{trott_autodock_2010}, we use the L-BFGS optimization method to generate poses that minimize this function.

The encoder, which we call the Twister, is composed of 32 TwistBlocks, each of which updates the internal representation of the ligand atoms and bonds, the protein binding pocket at both the residue and atomic level, and interactions between the ligand atoms with themselves, the protein residues, and the protein atoms. After every other TwistBlock, there is a residual connection; each of the hidden features is summed with the features from two layers earlier.

The Twister encoder takes as input the ligand molecular graph and a graph computed from the protein binding pocket. The nodes in the ligand graph are the heavy atoms, labeled with the element, formal charge, hybridization, number of bonded hydrogens, and aromaticity; the edges are the bonds, labeled with the bond order. Following previous studies \cite{jing_learning_2021, brocidiacono_bigbind_2022}, the nodes in the protein graph are the residues (labeled with the amino acid), and an edge exists between them if their \textalpha-carbons are within 10 \AA. The edges in this graph are annotated with the inter-\textalpha-carbon distances, encoded with Gaussian radial basis functions (RBFs). The model also requires all the pocket's heavy atoms, annotated with both the atom element and residue amino acid. The model initializes its learned representation to embeddings from all these input features; the ligand-ligand and ligand-receptor interaction features are initialized to zeros. These representations are then fed into the TwistBlocks.

The architecture of each TwistBlock is shown in Figure~\ref{overview}. We use the \verb|EGATConv| graph convolutions from DGL \cite{wang_deep_2020} to update the ligand atom and protein residue graph information. We use a variant of scaled-dot-product attention \cite{vaswani_attention_2017} to pass information between the residue-level and atomic-level representations of the protein pocket. We use this same attention mechanism to pass information to and from all the interaction representations. After all the graph convolutions and attention layers, we use dense layers followed by layer normalization \cite{ba_layer_2016} to finalize the update for each feature. We use a \verb|LeakyReLU| activation after each operation outlined above.

We use the Twister encoder to produce coefficients $C_{ijk}$ for each ligand-ligand and ligand-protein atom pair. We can then use these coefficients to score any ligand pose. For a given pose, we compute all the relevant inter-atomic distances $d_{ij}$, and use the following equation to predict the distance of each ligand atom $i$ to its true position ($\hat{D}_i$).
\begin{equation}
    \hat{D}_i = \sum_j \sum_k C_{ijk} e^{-\left(d_{ij} - \tilde{d}_k \right)^2  / \sigma^2 } + \beta
\end{equation}
Here $\beta$ is a learnable bias, $\tilde{d}_k$ are 24 RBF reference distances, evenly spaced from 0 to 32 \AA, and $\sigma$ is equal to 32/24. $i$ indexes over ligand atoms, $j$ indexes over both ligand and protein atoms, and $k$ indexes over the RBFs.

When training, we predict the individual ligand atom distances $\hat{D}_i$ directly. When inferring, we use the mean predicted distance  $\frac{1}{L}\sum_i \hat{D}_i$ as the global score for a given pose that we seek to minimize during inference.

\subsection{Training}

Following DiffDock, our training procedure consists of taking the ground-truth ligand pose of each complex and adding increasingly large amounts of noise to its translation, 3D rigid-body rotation, and torsional angles. We represent the 3D rotations as vectors pointed along the rotation axis whose magnitude is the rotation angle. 

We desire that, at the final timestep, the ligand is an essentially ``random'' conformation within the pocket. We use 16 timesteps; at each timestep, we add Gaussian noise to the translation vectors, rotation vectors, and torsional angles. We use a quadratic noise schedule with maximum standard deviations of 4 \AA\ for translation, 1.5 radians for the 3D rigid-body rotation, and 2.5 radians for each torsional angle.

Once we have the noised the ligand poses, we compute the distance of each atom to its position in the crystal structure. Our model predicts these distances, and we use the mean-squared-error (MSE) loss function when training. We note that, when training, our model is given the protein pocket from a different crystal structure than the ligand cognate structure (but aligned to the cognate structure so the ligand pose is reasonable). This is done to ensure our model performs well on cross-docking tasks.

\subsubsection{Dataset Preparation}

In order to follow the best practices for docking benchmarking outlined above, we start with the CrossDocked 2022 dataset \cite{francoeur_three-dimensional_2020}. This dataset already contains ligand crystal poses aligned to the poses of cognate receptors, so we can test cross-docking performance. We use data splits aligned with those from the CrossDocked paper so we can benchmark GNINA on complexes it wasn't trained on (see \ref{sec:running_gnina}). These splits were created by clustering the pockets in the dataset according to their ProBiS similarity score, a measure of structural similarity between two pockets \cite{konc_probis_2010}. Following \citet{brocidiacono_bigbind_2022}, the binding pockets are defined to be the set of all residues within 5 \AA\ of any ligand co-crystallized in that pocket. We filtered out all pockets with less than 5 residues or with bounding boxes greater than 42 \AA\ on any side.

\subsection{Inference}

To predict the ligand pose for a new protein-ligand complex, we start by running the Twister encoder to produce the score function coefficients from the pocket structure and 2D ligand graph. We then use RDKit \cite{noauthor_rdkit_nodate} to generate a 3D conformer and optimize it with the UFF force field \cite{rappe_uff_1992}. To generate a possible pose, we randomize the torsional angles of the conformer and place the ligand in the centroid of the binding site, rotated randomly and translated by Gaussian noise. The centroid of the binding site is defined as average coordinate of all the heavy atoms in the site. We then use the L-BFGS optimizer to compute translation, rotation, and torsional updates necessary to minimize the scoring function. We generate 16 poses with this method and return the one with the lowest final score.

\section{Results}

We benchmarked PLANTAIN on the CrossDocked test set and compared the performance with GNINA and Vina. Following prior convention, a pose is considered to be correct if it is within 2 \AA\ root-mean-square deviation (RMSD) of the correct pose. We also report the accuracy within 5 \AA\ RMSD. To account for the large disparity in the number of ligands per protein in the dataset, we report the average accuracy per pocket in the dataset (in addition to the total, unnormalized, accuracy).

In order to compare our method to DiffDock, we created a subset of the test set without any proteins from the DiffDock training set. We then run all the methods, including DiffDock, on this set. We note that DiffDock is disadvantaged in this benchmark because it is not given any knowledge of the binding pocket.

The results are shown in tables \ref{performance-all} and \ref{performance-diffdock}. PLANTAIN performs as well as GNINA, the previous state of the art, on 2 \AA\ accuracy, and beats it considerably in 5 \AA\ accuracy. Additionally, it runs ten times as fast. Intriguingly, PLANTAIN does not do as well at unnormalized 2 \AA\ accuracy (15.7\% vs 21.4\% normalized). This indicates that PLANTAIN struggled on some pockets in the dataset with large numbers of inhibitors, but performs well overall.

\begin{table}[t]
\caption{Performance of PLANTAIN, Vina, and GNINA on the CrossDocked test set. The accuracies reported are averaged over all the pockets in the dataset. The unnormalized accuracies are in parentheses.}
\label{performance-all}
\vskip 0.15in
\begin{center}
\begin{small}
\begin{sc}
\begin{tabular}{lcccr}
\toprule
Method & \% <2 \AA & \% <5 \AA\ & Runtime (s) \\
\midrule
Vina & 11.8 (14.4) & 39.1 (36.9) & 70.9 \\
GNINA & \textbf{21.5} (\textbf{21.5}) & 50.0 (46.1) & 51.2 \\
\textbf{PLANTAIN} & 21.4 (15.1) & \textbf{68.0} (\textbf{59.5}) & \textbf{4.9} \\
\bottomrule
\end{tabular}
\end{sc}
\end{small}
\end{center}
\vskip -0.1in
\end{table}

\begin{table}[t]
\caption{Performance of all models on the test set without the DiffDock training proteins. }
\label{performance-diffdock}
\vskip 0.15in
\begin{center}
\begin{small}
\begin{sc}
\begin{tabular}{lcccr}
\toprule
Method & \% <2 \AA\ & \% <5 \AA\ & Runtime (s) \\
\midrule
Vina & 11.7 (15.6) & 40.2 (37.9) & 73.7 \\
GNINA & 21.5 (\textbf{23.5}) & 51.7 (47.3) & 51.6 \\
DiffDock & 17.3 (11.6) & 46.6 (35.4) & 98.7 \\
\textbf{PLANTAIN} & \textbf{24.4} (15.2) & \textbf{73.7} (\textbf{71.9}) & \textbf{4.9} \\
\bottomrule
\end{tabular}
\end{sc}
\end{small}
\end{center}
\vskip -0.1in
\end{table}

\section{Discussion}

In this paper, we present PLANTAIN, a docking method that uses a novel neural network encoder to generate coefficients for a very fast scoring function. This scoring function is used to optimize an initially random ligand conformation. We show, using best practices for benchmarking, that our method achieves state-of-the-art accuracy while running significantly faster than competing methods.

There are many future directions to pursue. Most importantly, we desire to see how much PLANTAIN is able to assist in virtual screening. We plan on combining PLANTAIN pose prediction with GNINA's binding affinity prediction in order to test its effect on virtual screening benchmarks.

Additionally, the speed of PLANTAIN opens up many new possibilities. Previously, researchers have avoided flexible docking (wherein some or all of the protein residues are also able to move) because of performance concerns. It remains to be seen if PLANTAIN's fast scoring method is amenable to flexible docking.

Overall, PLANTAIN improves the state of the art in pose prediction, and there are clear directions for future improvement. We hope that this method will be useful in future drug discovery scenarios. Full source code for training and evaluating PLANTAIN is available at \url{https://github.com/molecularmodelinglab/plantain}.

\section{Acknowledgements}

The authors thank Rishal Aggarwal, Henry Dieckhaus, James Wellnitz, Josh Hochuli, Kathryn Kirchoff, and Travis Maxfield for their support and insightful discussions. We also thank Jack Lynch for his insights and GPUs. Studies reported in this paper were supported by the NIH grant R01GM140154.

\bibliography{references}
\bibliographystyle{icml2023}

\newpage
\appendix
\onecolumn

\section{Appendix}

\subsection{Training details}

PLANTAIN was trained for 67 epochs with a learning rate of $10^{-4}$ on a single NVIDIA GeForce GTX TITAN X GPU. We used the model saved on the 33rd epoch because it achieved the highest validation accuracy.

When training, we used a custom batch sampler to account for the fact that the protein-ligand complexes took up variable amounts of memory. Our sampler greedily adds complexes to the current batch until $B \cdot R  \geq 120 $, where $B$ is the proposed batch size and $R$ is the maximum number of receptor pocket residues in the batch.

\subsection{Benchmarking details}

\subsubsection{Running Vina}

In order to run AutoDock Vina, we prepared each ligand with Meeko \cite{durant_meeko_2022} and each receptor with OpenBabel \cite{oboyle_open_2011}. Following \citet{brocidiacono_bigbind_2022}, we used a bounding box that contained all the known ligands for that pocket with 4 \AA\ padding on all sides.

\subsubsection{Running Gnina}\label{sec:running_gnina}

We ran GNINA using the same bounding boxes as Vina. We also used an ensemble of 5 models from \citet{francoeur_three-dimensional_2020} that were not trained on the CrossDocked set data (by default, GNINA uses an ensemble of models that were trained on the entire CrossDocked set).

\subsubsection{Calculating runtimes}

We timed all the models on a laptop with an NVIDIA GeForce RTX 2060 Mobile GPU, using a single CPU thread. We note that Vina did not utilize the GPU.

\end{document}